\documentclass[aps,preprint]{revtex4}
\usepackage{amsmath}
\usepackage{bm}
\usepackage{graphicx}

\newcommand{\vq}{\mathbf{q}}

\newcommand{\vvr}{\mathbf{r}}

\newcommand{\hr}{\mathbf{\hat r}}

\newcommand{\vcr}{\mathbf{R}}
\newcommand{\vcq}{\mathbf{Q}}

\newcommand{\hcr}{\mathbf{\hat R}}

\newcommand{\deriv}{\mathop\partial\nolimits}

\newcommand{\bmu}{\bar\mu}
\newcommand{\bnu}{\bar\nu}

\begin{document}

\title{Collective multipole expansions and the perturbation theory in the
  quantum three-body problem} 

\author{A.~V.~Meremianin}
\email{meremianin@phys.vsu.ru}
\affiliation{Department of Theoretical Physics, 
Voronezh State University, 394006, Voronezh,  Russia}

\date{\today}

\begin{abstract}
The perturbation theory with respect to the potential energy of three
particles is considered.
The first-order correction to the continuum wave function of three free
particles is derived.
It is shown that the use of the collective multipole expansion of the free
three-body Green function over the set of Wigner $D$-functions can reduce the
dimensionality of perturbative matrix elements from twelve to six.
The explicit expressions for the coefficients of the collective multipole
expansion of the free Green function are derived.
It is found that the $S$-wave multipole coefficient depends only upon three
variables instead of six as higher multipoles do.
The possible applications of the developed theory to the three-body molecular
break-up processes are discussed.
\end{abstract}


\maketitle


\section{Introduction}
\label{sec:introduction}

The study of angular distributions in processes of three-particle
fragmentation is an important source of information about the dynamics 
of many physical objects such as atoms and molecules.

Till recently, the experimental technique allowed one to analyze the angular
distributions only for charged particles resulting from the many-body
fragmentation.
In such situations, the structure of the angular distributions is determined
by the Coulomb force.
For example, this is realized in the process of two-electron single-photon
ionization of Helium atom which is quite well studied both experimentally and
theoretically \cite{briggs00}.

During the last decade, the progress of experimental technique has made it
possible to investigate the angular distributions of neutral fragments arising
in the processes of three-atomic molecular break-up.
Namely, in \cite{datz00:_diss_recomb_water_angl_distr} the dissociative
recombination of water ion into the neutral atomic fragments has been studied.
In a series of experiments
\cite{helm99:_h3_first,galster04:_h3_prl,galster_helm05:_h3_results} the
angular distributions in the predissociation of the triatomic hydrogen into
three hydrogen atoms were investigated.
Clearly, the analysis of angular distributions in molecular break-up
processes could gain significant insight into the nature of chemical forces.

In molecular physics the dynamics of the interacting atoms is usually
described within the framework of the Born-Oppenheimer approximation.
In this approach the interaction of ions is represented by the set of
potential energy surfaces (PES) which are, in fact, the mean field of the
electrons.
It is important that, generally, PES cannot be decomposed into a sum of binary
potentials (i.e. pairwise interactions).
Rather, it will contain the terms which ``entangle'' the coordinates of
all particles thus making the standard methods (such as Faddeev
approach \cite{Faddeev_eqs60}) inapplicable.
Thus, the development of an adequate theory of many-particle molecular
fragmentation is a complicated task.
Therefore, it is highly desirable to have simple yet physically meaningful
models of the molecular break-up.

In the break-up into the neutral fragments often the situation is realized
when the kinetic energy of fragments prevails over their interaction
potential.
This opens the possibility to apply the perturbation theory to the calculation
of the wave function of three-particle continuum.
In the presented paper the lowest-order perturbation theory with respect to
the potential energy was applied to the quantum three-body system.
Even in this simplest case the calculation of the matrix elements leads
to twelve-dimensional integrals.
However, it turns out that the dimensionality of these integrals can be
reduced from twelve to six by employing the technique of collective multipole
expansions.
This technique was used in
\cite{avm05:_rigid_d3h_jpb} where such an expansion was derived for the
product of two three-dimensional plane waves 
$\exp i(\vq_1 \cdot \vvr_1 + \vq_2 \cdot \vvr_2)$.
Note that the results presented in the paper allows one to simplify the
calculation of the matrix elements within the perturbative approximation even
if the potential contains three-body (i.g. non-binary) terms.

The paper is organized as follows.
In Sec.~\ref{sec:general-formalism} the general equations of the perturbation
theory are derived.
They are based on the expression for the Green function corresponding to a
system of three free particles.
The calculation of matrix elements involving the perturbative wave functions
is considered in Sec.~\ref{sec:6d-hypersph}.
In that section it is also demonstrated how the calculations can be simplified
by employing the collective multipole expansion of the free Green function
over the basis set of Wigner $D$-functions.
The coefficients of that expansion are calculated in
Secs.~\ref{sec:mult-coeff-abf}, \ref{sec:hypersph-mult-coefs} using two
different approaches.
In Sec.~\ref{sec:mult-coeff-abf} the expressions for multipole coefficients are
derived by calculating the overlap integral of the free Green function with
Wigner $D$-function.
The expansion of the free Green function over the set of six-dimensional
hyperspherical harmonics is considered in Sec.~\ref{sec:hypersph-mult-coefs}.
The derived results are discussed in Sec.~\ref{sec:conclusion} where some
concluding remarks are also given. 
Appendix~\ref{sec:integr-euler-angles} contains details of computations of the
integral with the free Green function over the Euler angles.
The summary of properties of the set of six-dimensional hyperspherical
harmonics labeled by the particle's individual angular momenta quantum numbers
is given in Appendix~\ref{sec:hypersph-decomp-c2}.

The convention $\hbar=1$ is used throughout the text.
Capital letters denote the six-dimensional vectors, e.g. 
$\vcr=(\vvr_1, \vvr_2)$, where $\vvr_1, \vvr_2$ are three-dimensional vectors,
$R=|\vcr|=\sqrt{r_1^2+r_2^2}$.


\section{The general formalism}
\label{sec:general-formalism}

\begin{figure}[ptbh]
\centering
\includegraphics[width=5cm]{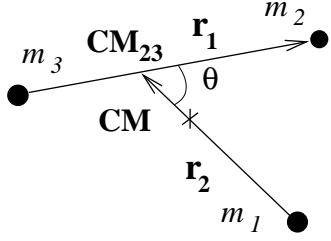}
\caption{Jacobi vectors for the three-body system. $CM_{23}$ is the CM of the
particles $m_{2}$ and $m_{3}$.}%
\label{fig:3bd-jacobi}%
\end{figure}

We begin by writing the Schroedinger equation for the system of three
particles with masses $m_1,m_2,m_3$,
\begin{equation}
\label{eq:schrod-3bd}
\left( - \frac{1}{2\mu_1}
\frac{\partial^{2}}{\partial\mathbf{r}_{1}^{2}}
- \frac{1}{2 \mu_2}
\frac{\partial^{2}}{\partial\mathbf{r}_{2}^{2}}
+U-E \right)  \Psi(\vvr_{1},\vvr_{2})=0.
\end{equation}
Here, $\vvr_{1}$ and $\vvr_2$ are Jacobi vectors (see
fig.~\ref{fig:3bd-jacobi}) and the reduced masses $\mu_1,\mu_2$ are defined by
\begin{equation}
\frac{1}{\mu_{1}}=\frac{1}{m_{2}}+\frac{1}{m_{3}},\quad\frac{1}{\mu_{2}}
  =\frac{1}{m_{1}}+\frac{1}{m_{2}+m_{3}}.
\label{eq:def-mu12}%
\end{equation}
It is convenient to replace $\vvr_1 \to \sqrt{\mu_1} \vvr_1$,
$\vvr_2 \to \sqrt{\mu_2} \vvr_2$ in eq. (\ref{eq:schrod-3bd}) which thereby
becomes
\begin{equation}
  \label{schrod-2}
\bigl( \Delta_6 -2 [U-E] \bigr)  \Psi(\vvr_{1},\vvr_{2})=0, 
\quad \Delta_6= \frac{\partial^{2}}{\partial\mathbf{r}_{1}^{2}} +
\frac{\partial^{2}}{\partial\mathbf{r}_{2}^{2}}.
\end{equation}
In this equation the potential energy depends on the masses of particles, but
the kinetic energy operator does not.

Below the situation is considered when the potential energy of the three-body
system is much smaller than the total (non-negative) energy $E$.
Introducing the wave number $Q$ of the continuum state by $E=Q^2/2$, we write
\begin{equation}
  \label{eq:pt-1}
  \left(\Delta_6 -2 U + Q^2 \right) \Psi (\vvr_1,\vvr_2) = 0,
\end{equation}
We approximate the total wave function by the combination 
$\Psi = \Psi^{(0)} + \Psi^{(1)}$, where $\Psi^{(0)}$ is independent of $U$,
and $\Psi^{(1)}$ scales as $\Psi \to c \Psi$ at $U \to c\,U$.
Substituting this decomposition into eq.~\eqref{eq:pt-1} and comparing the
coefficients which scale equally, we obtain
\begin{equation}
  \label{eq:pt-1a}
  \begin{split}
 \left(\Delta_6 + Q^2 \right) \Psi^{(0)} (\vvr_1,\vvr_2) &= 0, \\
\left(\Delta_6 + Q^2 \right) \Psi^{(1)} (\vvr_1,\vvr_2) &= 
2 U \,\Psi^{(0)}.
  \end{split}
\end{equation}
In the zeroth-order approximation the wave function is the product of two
three-dimensional plane waves
\begin{equation}
  \label{eq:pt-1b}
\Psi^{(0)} (\vvr_1,\vvr_2) = e^{i (\mathbf{Q}\cdot \vcr)}, \quad
(\mathbf{Q}\cdot \vcr)=(\vq_1 \cdot \vvr_1) + (\vq_2 \cdot \vvr_2),
\end{equation}
where $\vq_{1,2}$ are the linear momenta vectors conjugated to the Jacobi
vectors $\vvr_{1,2}$.
Consequently, the second equation in \eqref{eq:pt-1a} becomes
\begin{equation}
  \label{eq:pt-1c}
 \left(\Delta_6 + Q^2 \right) \Psi^{(1)} (\vvr_1,\vvr_2) = 
2 U \,  e^{i (\mathbf{Q}\cdot \vcr)}.
\end{equation}
In order to solve this equation we first consider its homogeneous form
\begin{equation}
  \label{eq:pt-2}
  \left(\Delta_6 + Q^2 \right) \Phi (\vcr) = 0.
\end{equation}
We are interested in a solution depending only on the hyperradius $R$. 
Therefore, we have to omit the angular part of the Laplacian $\Delta_6$,
which yields
\begin{equation}
  \label{eq:pt-3}
  \left(\frac{\partial^2 }{\partial R^2} + \frac{5}{R} 
\frac{\partial }{\partial R} +Q^2\right) \Phi(R) =0.
\end{equation}
Introducing the function $F= \Phi/R^2$, this equation transforms to
\begin{equation}
  \label{eq:pt-3a}
\frac{\partial^2 F}{\partial R^2} + \frac{1}{R}  
\frac{\partial F}{\partial R} + \left( Q^2 - \frac{4}{R} \right) F(R) =0.
\end{equation}
The general solution of this equation is the combination of two Hankel
functions $H^{(1)}_2 (QR)$ and $H^{(2)}_2 (QR)$ \cite{Bateman-II}.
We note that $H^{(1)}_2(QR)$ at large $R$ has the asymptote of the outgoing
spherical wave\footnote{The asymptote of the incoming spherical wave
  corresponds to Hankel function of the second kind $H^{(2)}_2(QR)$. This
  solution must be taken when considering the fragmentation processes
  \cite{landau3tom_eng}.}
\begin{equation}
  \label{eq:pt-3b}
 H^{(1)}_2 (QR) \approx - \left( \frac{2}{\pi QR} \right)^{1/2}
\,e^{i (QR-\pi/4)}, \quad R \to \infty.
\end{equation}
Thus, the solution of (\ref{eq:pt-3}) can be chosen to be
\begin{equation}
  \label{eq:pt-4}
  \Phi(\vcr) =\frac{H^{(1)}_2 (QR)}{R^2}.
\end{equation}
This function satisfies eq.~(\ref{eq:pt-2}) everywhere except $R=0$.
Below we shall prove that
\begin{equation}
  \label{eq:pt-gf}
  G(\vcr) = G(R) = -i \left( \frac{Q}{4\pi R}\,\right)^2
H^{(1)}_2 (QR).
\end{equation}
is the Green function of the equation \eqref{eq:pt-2}, that is
the solution of the inhomogeneous equation 
\begin{equation}
  \label{eq:pt-4a}
   \left(\Delta^{(6)} + Q^2 \right) G (\vcr) = \delta(\vcr). 
\end{equation}
Indeed, at $R > 0$ this function satisfies the homogeneous equation
\eqref{eq:pt-2}.
At the limit $R \to 0$ there is a divergency,
\begin{equation}
  \label{eq:pt-gf-1}
  G(\vcr) \approx -i \left( \frac{Q}{4\pi R}\,\right)^2
\frac{-i}{\pi}\, \left( \frac{2}{Q R}\,\right)^2
=\frac{-1}{4 \pi^3 R^4}.
\end{equation}
From the properties of Dirac $\delta$-function it follows that integral of
\eqref{eq:pt-4a} taken over the arbitrary region which includes the point
$\vcr=0$ must be equal to unity.
We calculate the six-dimensional integral of \eqref{eq:pt-4a} taken over the
infinitesimal sphere $V_\epsilon$ with its center in $\vcr=0$,
\begin{equation}
  \label{eq:pt-gf-2}
\int_{V_\epsilon}  \left(\Delta_6 + Q^2 \right) G(R) d^6 R 
= \int_{S_\epsilon} d \mathbf{S} \cdot \nabla_6  G(R) \Bigr|_{R=\epsilon}
+ \pi^3 Q^2 \int_0^\epsilon G(R) R^5 d R.
\end{equation}
Here, the Gauss theorem has been used in order to transform the volume
integral into the integral over the surface of $6$-sphere $S_\epsilon$ with
the radius $\epsilon$.
Noting the expression for the vector surface element
$d \mathbf{S} = (\vcr/R)\,\epsilon^5 d S$,  the integral \eqref{eq:pt-gf-2}
at $\epsilon \to 0$ evaluates to
\begin{equation}
  \label{eq:pt-gf-3}
  \int_{V_\epsilon}  \left(\Delta_6 + Q^2 \right) G(R) d^6 R 
\approx
\frac{1}{\pi^3}\, \int_{S_\epsilon} d S 
- \frac{Q^2}{4} \int_0^\epsilon R d R
= 1 - \frac{\epsilon^2 Q^2}{8} \to 1.
\end{equation}
As is seen, $G(\vcr)$ defined by eq.~(\ref{eq:pt-gf}) does satisfy the
equation \eqref{eq:pt-4a}.

Having Green function (\ref{eq:pt-gf}), the solution of the equation
\eqref{eq:pt-1c} can be immediately written as
\begin{equation}
  \label{eq:pt-psi}
 \Psi^{(1)} (\vcr) = 2 \int  e^{i (\mathbf{Q}\cdot \vcr')}\,
U(\vcr')\, G(| \vcr-\vcr'|)\, d^6 R'.
\end{equation}
Thus, the continuum wave function $\Psi(\vcr)$ of three particles within the
perturbative approach has the form
\begin{equation}
  \label{eq:pt-psi-tot}
\Psi(\vcr)=  \Psi_{\vcq} (\vcr) = e^{i\,(\mathbf{Q}\cdot\vcr)}
 -i \frac{Q^2}{8\pi^2}\, 
\int  e^{i (\mathbf{Q}\cdot \vcr')} \, U(\vcr') \,
\frac{H^{(1)}_2 (Q| \vcr-\vcr'|)}{| \vcr-\vcr'|^2} \, d^6 R'.
\end{equation}


\section{The calculation of the matrix elements with
the correction term} 
\label{sec:6d-hypersph}

In many physical applications it is necessary to calculate the matrix
element 
\begin{equation}
  \label{eq:me-1}
M_{j\mu}=  \langle \Psi_{j\mu} | \mathcal{O} | \Psi_{\vcq}  \rangle,
\end{equation}
where $\mathcal{O}$ is the transition operator and $\Psi_{j\mu}$ is the
wave function which corresponds to the state having the total angular momentum
numbers $j\mu$.
Without loss of generality, we can assume that $\mathcal{O}$ is a scalar
operator.
This situation realizes, for example, in the case of the three-atomic
molecular predissociation where $\mathcal{O}$ represents the operator of
non-adiabatic couplings \cite{galster_helm05:_h3_results}.
If the transition operator is a tensor $\mathcal{O}_{lm}$ then the product 
$\Psi_{j\mu} \mathcal{O}_{lm}$ can be decomposed into irreducible parts
\cite{Varsh} so that the matrix element will be decomposed into the sum of
matrix elements of the kind (\ref{eq:me-1}).

We note that \eqref{eq:me-1} is the six-dimensional integral,
\begin{equation}
  \label{eq:me-2}
  M_{j\mu} = \int \Psi^*_{j\mu}(\vcr)\, \mathcal{O} \, \Psi_{\vcq} (\vcr)\,
d^6 R.
\end{equation}
Among the six variables three can be chosen to be the collective angles
determining the orientation of the whole system in space.
The remaining three ``shape'' variables $\xi$ determine the internal dynamics
of the system and they can be chosen to be $\xi=r_1,r_2,\cos\theta$,
where $\theta$ is the angle between $\vvr_1$ and $\vvr_2$.
Thus, the transition operator $\mathcal{O}$ is the function of $\xi$, 
$\mathcal{O}=\mathcal{O}(\xi)$.

Substituting the decomposition (\ref{eq:pt-psi-tot}) into (\ref{eq:me-2}) one
arrives at the expression
\begin{equation}
  \label{eq:me-2a}
  M_{j\mu} = M^{(0)}_{j\mu} + M^{(1)}_{j\mu},
\end{equation}
where the zero- and first-order matrix elements are defined by
\begin{equation}
  \label{eq:me-2b}
  \begin{split}
  M^{(0)}_{j\mu} =& \int \Psi^*_{j\mu}(\vcr)\, \mathcal{O} \,
e^{i (\mathbf{Q}\cdot \vcr')}\, d^6 R, \\
  M^{(1)}_{j\mu} =&  -i \frac{Q^2}{8\pi^2}\,
\int \Psi^*_{j\mu}(\vcr)\, \mathcal{O} \,
\int  e^{i (\mathbf{Q}\cdot \vcr')} \, U(\xi') \,
\frac{H^{(1)}_2 (Q|\vcr'-\vcr|)}{|\vcr'-\vcr|^2} \, d^6 R'\, d^6 R.
  \end{split}
\end{equation}
Here, by writing $U(\xi')$ we make the assumption that the potential energy
depends only on the three shape variables.
This is true if the three-body system in not a subject of external forces.

The calculation of the matrix element $M^{(0)}_{j\mu}$ has been considered in
\cite{avm05:_rigid_d3h_jpb}.
Therefore, below we concentrate on the problem of calculation of the
first-order (with respect to the potential) matrix element $M^{(1)}_{j\mu}$.

As is seen, $M^{(1)}_{j\mu}$ is determined by the 12-dimensional integral.
However, six of twelve variables are the collective angles describing the
rotation of the whole system from the coordinate frame defined by Jacobi
vectors $(\vvr_1,\vvr_2)$ to the frame defined by $(\vvr'_1, \vvr'_2)$ and
from the frame $(\vvr_1, \vvr_2)$ to $(\vq_1, \vq_2)$.
Noting that the potential energy does not depend on collective angles, one can
try to integrate them out analytically.
In order to do so one has to expand the integrands over the basis of angular
functions depending on the collective angles.
It is convenient to choose as the angular basis the set of Wigner D-functions
\cite{Varsh}.

The initial state wave function $\Psi_{j \mu}(\xi)$ can be decomposed into the
combination of $(2j+1)$ ``internal'' wave functions $\psi_{j\nu}(\xi)$,
depending on three shape variables $\xi$,
\begin{equation}
  \label{eq:psi-i1}
  \Psi_{j \mu}(\vcr) = \sum_{\nu=-j}^j \psi_{j\nu} (\xi)\,
  D^j_{\nu,\mu}(\Omega),
\end{equation}
where $\Omega$ denotes three collective Euler angles describing the rotation
from the body-fixed frame (BF) defined by the Jacobi vectors 
$\vcr=(\vvr_1, \vvr_2)$ to the laboratory frame (LF) defined by the momenta
vectors $\vcq=(\vq_1, \vcq_2)$.

At this stage we have to evaluate the angular integral
\begin{equation}
  \label{eq:aint-1}
  I_{\vcq} (\vcr) = \int e^{i (\mathbf{Q}\cdot \vcr')} \,
\frac{H^{(1)}_2 (Q|\vcr'-\vcr|)}{|\vcr'-\vcr|^2} \, d^3 \Omega',
\end{equation}
where $\Omega'$ denotes three Euler angles describing the rotation from
$\mathrm{BF}^\prime$ defined by the pair of Jacobi vectors
$\vcr'=(\vvr'_1, \vvr'_2)$ to LF.

The calculation of $I_{\vcq} (\vcr)$ can be performed by taking the multipole
expansion of both integrand functions,
\begin{equation}
  \label{eq:i-1}
  \begin{split}
    e^{i (\mathbf{Q}\cdot \vcr')} &= \sum_{j=0}^\infty \sum_{\mu,\nu=-j}^j
    F^{(j)}_{\mu,\nu} (\xi_q;\xi') \, D^j_{\mu,\nu}(\Omega'), \\
\frac{H^{(1)}_2 (Q|\vcr'-\vcr|)}{|\vcr'-\vcr|^2} &= 
\sum_{j=0}^\infty \frac{2j+1}{8\pi^2}
 \sum_{\mu,\nu=-j}^j
 G^{(j)}_{\mu,\nu} (\xi; \xi') \, D^j_{\mu,\nu}(\Omega''),
  \end{split}
\end{equation}
where $\Omega''$ describes the rotation from BF defined by 
$(\vvr_1, \vvr_2)$ to $\mathrm{BF}^\prime$ defined by $(\vvr'_1, \vvr'_2)$.
In the upper equation $\xi_q$ denotes three ``shape'' variables in the
momentum space, e.g. $\xi_q=q_1,q_2,\cos\chi$, where $\chi$ is the angle
between $\vq_1$ and $\vq_2$.

Note that the rotation $\Omega''$ can be presented as a product of rotations
\begin{equation}
  \label{eq:om-1}
  \Omega'' = \Omega \times {\Omega'}^{-1}.
\end{equation}
In terms of Wigner D-functions this equation reads \cite{Varsh},
\begin{equation}
  \label{eq:om-2}
 D^j_{\mu,\nu}(\Omega'') = \sum_{\nu'=-j}^j
D^j_{\mu,\nu'}(\Omega)\,  D^j_{\nu',\nu}({\Omega'}^{-1})
= \sum_{\nu'=-j}^j
D^j_{\mu,\nu'}(\Omega)\,  [D^j_{-\nu,-\nu'}(\Omega')]^*.
\end{equation}
Substituting this equation and eqs.~(\ref{eq:i-1}) into eq.~(\ref{eq:aint-1}),
we arrive at the identity
\begin{multline}
 \label{eq:aint-2}
  I_{\vcq} (\vcr) = \sum_{j,j'=0}^\infty \frac{2j'+1}{8\pi^2}
\sum_{\mu,\mu',\nu,\nu',\nu''}
 F^{(j)}_{\mu,\nu} (\xi_q,\xi') \,  G^{(j')}_{\mu',\nu'} (\xi; \xi') \,
D^{j'}_{\mu',\nu''}(\Omega) \\
\times
 \int  D^j_{\mu,\nu}(\Omega')\, [D^{j'}_{-\nu',-\nu''}(\Omega')]^*\,
 d \Omega'.
\end{multline}
The integration gives 
$\delta_{j,j'} \delta_{\mu,-\nu'} \delta_{\nu,-\nu''}\,8\pi^2/(2j+1)$ so that
\begin{equation}
  \label{eq:aint-3}
  I_{\vcq} (\vcr) = \sum_{j=0}^\infty \sum_{\mu,\mu',\nu}
 F^{(j)}_{\mu',-\nu} (\xi_q,\xi') \,  G^{(j)}_{\mu,-\mu'} (\xi; \xi') \,
D^{j}_{\mu,\nu}(\Omega).  
\end{equation}
Inserting this equation together with eq.~(\ref{eq:psi-i1}) for the initial
wave function into eq.~(\ref{eq:me-2b}) for the matrix element $M^{(1)}_{j\mu}$,
one can integrate out the angles $\Omega$.
The result is
\begin{equation}
  \label{eq:m1-1}
  M^{(1)}_{j\mu} = \frac{ -i\, Q^2}{2j+1}
\sum_{\nu=-j}^j
\int \psi^*_{j\nu}(\xi)\, \mathcal{O}(\xi) 
\sum_{\mu'=-j}^j
\int G^{(j)}_{\nu,-\mu'} (\xi; \xi') \, F^{(j)}_{\mu',-\mu} (\xi_q,\xi') \,
U(\xi')\,
 d^3 \xi'\, d^3 \xi.
\end{equation}
As is seen, the 12-dimensional integral (\ref{eq:me-2b}) is reduced to the
6-dimensional integral.

In the next sections we consider the problem of the calculation of the
collective multipole coefficients $G^{(j)}_{\mu,\nu}(\xi;\xi')$ introduced in
eq.~(\ref{eq:i-1}).
As a preliminary, we note that these coefficients obey the symmetry relation
\begin{equation}
  \label{eq:gjm-sym}
  [ G^{(j)}_{\mu,\nu} (\xi;\xi') ]^* = (-1)^{\mu-\nu}\,
G^{(j)}_{-\mu,-\nu}(\xi;\xi')
\end{equation}
which follows from the symmetry property of D-functions 
$D^j_{\mu,\nu}(\Omega) = (-1)^{\mu-\nu} [D^j_{-\mu,-\nu}(\Omega)]^*$.
Thus, in practice it is enough to calculate only the coefficients 
with $\mu \ge \nu$.

For the sake of completeness we present also the expression for the volume
element of the shape space for two most often used choices of the
set of shape variables $\xi$
\begin{equation}
  \label{eq:vel-1}
  \int d^3 \xi = \int_0^\infty r_1^2 dr_1 \int_0^\infty r^2_2 d r_2\,
\int_0^\pi \sin\theta \,d\theta
= \int_0^\infty R^5\, dR \int_0^{\pi/2} \frac{\sin^2 2\alpha}{4} d \alpha
\int_0^\pi \sin\theta \,d\theta,
\end{equation}
where the second equation corresponds to the hyperspherical set 
$\xi=R,\alpha,\theta$, where $R=\sqrt{r^2_1+r^2_2}$ is the hyperradius and 
$\alpha=\arctan(r_2/r_1)$ is the hyperangle, so that
\begin{equation}
  \label{eq:def-hsc}
  r_1 = R\, \cos\alpha,\quad r_2 = R\, \sin\alpha, \quad
0 \le \alpha \le \pi/2.
\end{equation}


\section{Multipole coefficients for the angle-bisector gauge}
\label{sec:mult-coeff-abf}

Using the orthogonality of $D$-functions the expression for the multipole
coefficients $G^{(j)}_{\mu,\nu}$ can be written as the three-dimensional
integral,
\begin{equation}
  \label{eq:cg-1}
  G^{(j)}_{\mu,\nu} (\xi; \xi') = \int 
\frac{H^{(1)}_2 (Q|\vcr'-\vcr|)}{|\vcr'-\vcr|^2}
[D^j_{\mu,\nu}(\Omega'')]^* \, d^3 \Omega''.
\end{equation}
Here, apart of D-functions, Euler angles $\Omega''$ enter the integrand
through the term $| \vcr - \vcr'|$ which is
\begin{equation}
  \label{eq:r-r}
  | \vcr - \vcr'| = \sqrt{R^2 + {R'}^2 - 2 p}, \quad
p= (\vvr_1 \cdot \vvr'_1) + (\vvr_2 \cdot \vvr'_2),
\end{equation}
where only the parameter $p$ depends on $\Omega''$.

At this stage, one has to specify explicitly how the axes of the BF and
$\mathrm{BF}'$ are connected to the Jacobi vectors $(\vvr_1, \vvr_2)$ and 
$(\vvr'_1, \vvr'_2)$.
This procedure is not unique and, therefore, it can be denoted as
the ``gauge convention'' \cite{littlejohn97:_gauge}.

\begin{figure}
  \centering
  \includegraphics[width=4.5cm]{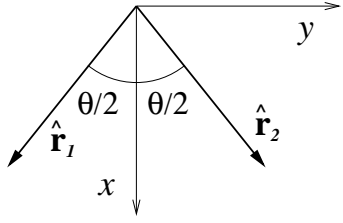}
  \caption{The angle-bisector gauge:
  the $x$-axis is directed along the bisector of the angle between the unit
  vectors $\hr_1$ and $\hr_2$. The $z$-axis is directed along the vector
  product $[\hr_1 \times \hr_2]$.
  $\mathrm{BF}'$ is defined analogously. }
  \label{fig:abf}
\end{figure}

Below we use angle-bisector gauge for both BF and $\mathrm{BF}'$, see
fig.~\ref{fig:abf}.
Since $p$ is scalar, it does not matter in which frame we calculate it.
Let it be $\mathrm{BF}'$.
In this frame the components of $\vcr'$ have the form
\begin{equation}
  \begin{split}
\vvr'_{1}=& r'_1\,
  [\cos (\theta'/2), -\sin (\theta'/2), 0], \\
\vvr'_{2}=& r'_2\,
  [\cos (\theta'/2), \sin(\theta'/2), 0 ],
  \end{split}
\label{eq:def-abf}%
\end{equation}
The components of $\vcr$ in $\mathrm{BF}'$ can be calculated
using the equation
\begin{equation}
({\tilde \vvr}_n)_{i} =\sum_{k=x,y,z} (\vvr_n)_{k}\,
\mathcal{A}_{ki} (\Omega''),
\quad n=1,2.
\label{eq:def-a_ik}%
\end{equation}
where $\mathcal{A}_{ki} (\Omega'')$ denotes the components of the rotation
matrix \cite{Varsh} and $(\vvr_n)_k$ is $k$-th component of $\vvr_n$ in BF
where it is given by eq.~(\ref{eq:def-abf}) with dashes removed.
From eq.~(\ref{eq:def-a_ik}) follows the identity
$p=\sum_{i,k}( (\vvr'_1)_i\,({\tilde \vvr}_1)_{k}\mathcal{A}_{ki}
+(\vvr'_2)_i\,({\tilde \vvr}_2)_{k}\mathcal{A}_{ki} )$.

Using explicit form of the rotation matrix $\mathcal{A}$,
after some algebraic transformations described in detail in
\cite{avm05:_rigid_d3h_jpb}, the parameter $p$ can be written as
\begin{equation}
  \label{eq:def-p}
  p = a_1\, \sin(\alpha+\gamma+\delta_1) -
a_2\, \sin(\alpha-\gamma+\delta_2).
\end{equation}
where the parameters are defined by
\begin{equation}
  \label{eq:def-a}
  \begin{split}
    a_1 &= \rho_1 \left(\cos\frac{\beta}{2}\right)^2, \quad
a_1 = \rho_2 \left(\sin\frac{\beta}{2}\right)^2, \\
\rho_{1,2}^{2} &= (r_1 r'_1)^{2} + (r_2 r'_2)^{2} +
    2 r_1 r_2 r'_1 r'_2 \,\cos(\theta \mp \theta'),
  \end{split}
\end{equation}
where $\rho_1$ corresponds to $(\theta-\theta')$.
The additional phases $\delta_{1,2}$ are defined by
\begin{equation}
\begin{split}
\sin\delta_{1,2} = & 
\frac{r_{1}r'_{1}+r'_{2}r_{2}}{\rho_{1,2}}\,
\cos\frac{\theta \mp \theta'}{2}, \\
\cos\delta_{1,2} =&
\frac{r_{2}r'_{2}-r'_{1}r_{1}}{\rho_{1,2}}\,
\sin\frac{\theta \mp \theta'}{2},
\end{split}
\end{equation}
where the upper sign corresponds to $\delta_1$.

In order to calculate the integral in (\ref{eq:cg-1}) we employ the series
expansion 
(eq.~(7.15.18) of \cite{Bateman-II}),
\begin{equation}
  \label{eq:h-1}
  \frac{H^{(1)}_2 \left( Q | \vcr'-\vcr | \right)  }{|\vcr'-\vcr|^2}
= \sum_{n=0}^\infty \frac{( Q\,p)^{n}}{n!} 
\frac{H^{(1)}_{n+2} \left( Q \sqrt{R^2+R'^2} \right)}{(R^2+R'^2)^{n/2+1} }.
\end{equation}
Thus, the expression for the collective multipoles can be written as
\begin{equation}
  \label{eq:g-1}
  G^j_{\mu,\nu} (\xi,\xi') = 
\sum_{n=0}^\infty P^{(j\mu\nu)}_n \, Q^{n} 
\frac{H^{(1)}_{2m+\mu+2} \left( Q \sqrt{R^2+R'^2} \right)}{(R^2+R'^2)^{n/2+1}},
\end{equation}
where the functions $P^{(j\mu\nu)}_n$ are defined by the integral
\begin{equation}
  \label{eq:i-4}
P^{(j\mu\nu)}_n = \frac{1}{n!}\,
 \int p^n \, [D^j_{\mu,\nu} (\Omega'')]^* \, d^3 \Omega''.
\end{equation}
This integral is calculated in Appendix~\ref{sec:integr-euler-angles} (see
eq.~(\ref{eq:int-pn})).
In Appendix~\ref{sec:integr-euler-angles} it is also demonstrated that the
functions $P^{(j\mu\nu)}_n$ are non-zero only for indices $\mu,\nu$ and $n$
having the same parity. 
Hence, the multipoles $G^j_{\mu,\nu}$ vanish if $\mu$ and $\nu$ have different
parity.

Below we present explicit expressions for the parameter $G^j_{\mu,\nu}$ for
the most simple cases $j=0,1$.
For $S$-state the parameter $P^{(000)}_n$ is given by eq.~(\ref{eq:g000})
which leads to
\begin{equation}
  \label{eq:g-s}
  G^0_{0,0} (\xi,\xi') = \frac{8 \pi^2}{R^2+{R'}^2}\, 
\sum_{n=0}^\infty \frac{1}{(2n+1)!}
\left(
\frac{\rho_1 \rho_2 \, Q^2}{R^2+R'^2}
\right)^{n}
P_n \left( w \right)
H^{(1)}_{2n+2} \left( Q \sqrt{R^2+R'^2} \right),
\end{equation}
where $w= (\rho_1^2 + \rho_2^2)/(2 \rho_1 \rho_2)$ and $P_n(w)$ denotes the
Legendre polynomial.
For $P$-state ($j=1$) there are three multipoles:
$G^1_{0,0}$, $G^1_{1,1}=-(G^1_{-1,-1})^*$, $G^1_{1,-1}=(G^1_{-1,1})^*$.
These coefficients correspond to the states with different spatial parity
\cite{meremianin03:_phys_rep}.
Namely, $G^1_{0,0}$ is the pseudotensor parity multipole (state $P^{even}$), and
$G^1_{1,\pm1}$ are polar tensor multipoles (state $P^{odd}$).
We present here only the expression for $G^1_{0,0}$,
\begin{multline}
  \label{eq:g-p}
  G^1_{0,0} (\xi,\xi') = \frac{8 \pi^2}{R^2+{R'}^2}\, 
\left(
\frac{\rho^2_1 - \rho^2_2}{\rho_1 \rho_2}
\right)
\sum_{n=0}^\infty \frac{1}{(2n+2)!}
\left(
\frac{\rho_1 \rho_2 \, Q^2}{R^2+R'^2}
\right)^{n}
\, P'_n(w) \\
\times
H^{(1)}_{2n+2} \left( Q \sqrt{R^2+R'^2} \right),
\end{multline}
where $P'_n (w) = d P_n(w)/ d w$.


\section{The hyperspherical expansion of the multipole 
coefficients}
\label{sec:hypersph-mult-coefs}

The hyperspherical form of the multipole coefficients can be derived based on
the following expansion of the three-body free Green function \cite{Bateman-II}
\begin{equation}
  \label{eq:pt-5}
\frac{  H^{(1)}_2 (Q|\vcr'-\vcr|)}{|\vcr'-\vcr|^2} 
= \frac{4}{(QRR')^2}
\sum_{n=0}^\infty (n+2)\,C^2_n (\hcr \cdot \hcr')\,
J_{n+2}(QR)\,H^{(1)}_{n+2}(QR'), \quad R < R',
\end{equation}
where $\hcr \cdot \hcr' = p/(RR')$ and $p$ is defined in (\ref{eq:r-r});
$C^2_n(\hcr \cdot \hcr')$ is $n$-th order Gegenbauer polynomial
\cite{Bateman-II}.
If $R > R'$ then the substitution $R \leftrightarrow R'$ must be made in 
(\ref{eq:pt-5}).

According the definition (\ref{eq:cg-1}), the expression for the multipole
coefficients can be written as
\begin{equation}
  \label{eq:g-hsh}
  G^j_{\mu,\nu} (\xi;\xi') = \frac{1}{(QRR')^2}
\sum_{n=0}^\infty h_n^{(j\mu\nu)} \,J_{n+2}(QR)\,H^{(1)}_{n+2}(QR'), 
\quad R < R',
\end{equation}
where $J_{n+2}(QR)$ is Bessel function and the functions $h^{(j\mu\nu)}_n$
are defined by the integral
\begin{equation}
  \label{eq:int-hsh}
h^{(j\mu\nu)}_n = 4 (n+2)\,
\int C^2_n (\hcr \cdot \hcr')\, [D^j_{\mu,\nu}(\Omega'')]^* \, d^3 \Omega''.
\end{equation}
In order to calculate this integral we note that the Gegenbauer polynomial in
(\ref{eq:pt-5}) is proportional to the scalar product of six-dimensional
hyperspherical harmonics \cite{Bateman-II,avery85jmp:_hyp_spher},
\begin{equation}
  \label{eq:def-y6d-4}
C^2_n (\hcr \cdot \hcr') = \frac{\pi^2}{2\,(n+2)}\,
(Y_n (\hcr) \cdot Y_n (\hcr') ).
\end{equation}
It is convenient to define harmonics $Y_{n}$ to be the eigenfunctions of the
angular momentum operators corresponding to the Jacobi vectors $\vvr_1$ and
$\vvr_2$.
In this approach, six-dimensional spherical harmonics are proportional to the
product of conventional three-dimensional harmonics depending on the spherical
angles of unit vectors $\hr_1$ and $\hr_2$,
\begin{equation}
  \label{eq:def-y6d}
  Y_{nqlmm'}(\hcr ) = C^{(n-2q-l,\,l)}_q (\alpha)\, Y_{lm} (\hr_1)\,
Y_{(n-2q-l)m'} (\hr_2),
\end{equation}
where the functions $C^{(n-2q-l,\,l)}_q (\alpha)$ depending on the hyperangle
$\alpha$ are defined by eq.~(\ref{eq:def-cpol}) of
Appendix~\ref{sec:hypersph-decomp-c2}.

The expression (\ref{eq:def-y6d-3}) for the scalar product of six-dimensional
harmonics contains the product of two Legendre polynomials depending on
$(\hr_1 \cdot \hr'_1)$ and $(\hr_2 \cdot \hr'_2)$.
Thus, in order to calculate the integral (\ref{eq:int-hsh}) it is necessary to
decompose the product of two Legendre polynomials into a combination of
angular functions corresponding to the states with well-defined values of the
total angular momentum $j$.
This can be done by expressing the Legendre polynomials via three-dimensional
spherical harmonics.
Next, the product of spherical harmonics can be re-written in terms of bipolar
harmonics,
\begin{equation}
  \label{eq:lp-2}
P_l (\hr_1 \cdot \hr'_1) \,   P_{l'} (\hr_2 \cdot \hr'_2) 
= \sum_{j=|l-l'|}^{l+l'} (-1)^{l+l'-j} 
\left( C^{ll'}_{j} (\hr_1, \hr_2) \cdot C^{ll'}_{j} (\hr'_1, \hr'_2)\right),
\end{equation}
where the bipolar harmonics $C^{ll'}_{j} (\hr_1, \hr_2)$ are tensor product of
two spherical harmonics \cite{Varsh},
\begin{equation}
  \label{eq:def-bh}
  C^{ll'}_{j\mu} (\hr_1, \hr_2) = \sum_{m, m'} C^{j\mu}_{lm\,l'm'}
  C_{lm}(\hr_1)\, C_{l'm'} (\hr_2).
\end{equation}
Here, $C_{lm} = \sqrt{4\pi/(2l+1)}\, Y_{lm}$ are modified spherical harmonics
and $C^{j\mu}_{lm\,l'm'}$ are Clebsch-Gordan coefficients.
The scalar product of bipolar harmonics does not depend on the
choice of the coordinate frame in which it is calculated.
Let this frame be $\mathrm{BF}'$, then
\begin{multline}
  \label{eq:bh-bf}
\left( C^{ll'}_{j} (\hr_1, \hr_2) \cdot C^{ll'}_{j} (\hr'_1, \hr'_2)\right)
= \sum_{\nu}  (-1)^\nu\,[ C^{ll'}_{j-\nu} (\hr_1, \hr_2) ]_{BF'}\,
[C^{ll'}_{j\nu} (\hr'_1, \hr'_2) ]_{BF'} \\
= \sum_{\mu,\nu}  (-1)^\nu\, [C^{ll'}_{j-\nu} (\hr'_1, \hr'_2) ]_{BF'} \,
[C^{ll'}_{j\mu} (\hr_1, \hr_2)]_{BF} \, D^j_{\mu,\nu} (\Omega''),
\end{multline}
where the subscript BF ($\mathrm{BF}'$) denotes the coordinate frame in
which the components must be calculated.
The second identity in (\ref{eq:bh-bf}) follows from the tensor transformation
rule under the rotation of the coordinate frame \cite{Varsh}.
Now the integral (\ref{eq:int-hsh}) can be easily calculated using the
orthogonality of $D$-functions,
\begin{multline}
  \label{eq:bh-h}
h^{(j\mu\nu)}_n =
(-1)^{n-j+\nu}\,\frac{4 \pi^3}{2j+1}
\sum_{q,l=0} (2l+1)\,(2[n-2q-l]+1)\,
C^{(n-2q-l,\,l)}_q (\alpha)\,C^{(n-2q-l,\,l)}_q (\alpha') \\
\times
 [C^{l\, (n-2q-l)}_{j-\nu} (\hr'_1, \hr'_2) ]_{BF'} \,
[C^{l\,(n-2q-l)}_{j\mu} (\hr_1, \hr_2)]_{BF},
\end{multline}
where the indices $q,l$ take non-negative integer values so that 
$n-2q-l \ge 0$.
To derive the explicit form of $h^{(j\mu\nu)}_n$ it is necessary to
specify the gauge convention.
The only exception is the case of $j=0$ ($S$-wave part of the expansion
(\ref{eq:pt-5})), when the coefficient $h^{(000)}_n$ is scalar
\begin{equation}
  \label{eq:bh-s-1}
 h^{(000)}_{2n} = 4 \pi^3
\sum_{l=0}^n (2l+1)\,
C^{(l,l)}_{n-l} (\alpha)\,C^{(l,l)}_{n-l} (\alpha')\,
P_l (\cos\theta)\, P_l (\cos\theta').
\end{equation}
Note that the coefficients $h^{(000)}_n$ with odd $n$ vanish, 
$h^{(000)}_{2n+1}=0$.

For $j>0$ we use the angle-bisector gauge (see fig.~\ref{fig:abf} above).
In that gauge, the components of bipolar harmonics in (\ref{eq:bh-h}) have the
form
\begin{multline}
  \label{eq:bh-abf}
[C^{l\,(n-2q-l)}_{j\mu} (\hr_1, \hr_2)]_{BF}
= C^{l\,(n-2q-l)}_{j\mu} (\pi/2, -\theta/2, \pi/2, \theta/2) \\
= \sum_{m}
C^{j\mu}_{lm\,(n-2q-l) (\mu-m)}
C_{lm}(\pi/2, -\theta/2)\, C_{(n-2q-l)\, (\mu-m)} (\pi/2, \theta/2).
\end{multline}
Note that the explicit expression for the spherical harmonics
$C_{lm}(\pi/2,\theta)$ is rather simple \cite{Varsh}
\begin{equation}
  \label{eq:clm-half-pi}
  C_{lm} \left( \frac{\pi}{2}, -\frac{\theta}{2} \right) = 
  \begin{cases}
    (-1)^{(l+m)/2}\,e^{-im\theta/2}\,
       \sqrt\frac{(l+m-1)!!\,(l-m-1)!!}{(l+m)!!\,(l-m)!!},
     \quad l+m=\mathrm{even} \\
    0, \quad l+m=\mathrm{odd}.
  \end{cases}
\end{equation}
From this identity and the symmetry properties of Clebsch-Gordan
coefficients it follows that $h^{(j\mu\nu)}_n$ are non-zero only for $n$,
$\mu$ and $\nu$ having the same parity.

For $j=1$ and $\mu=\nu=0$ it is convenient to use the reduction formula for
bipolar harmonics \cite{Man-96},
\begin{equation}
  \label{eq:bh-j1}
[  C^{ll}_{10} (\hr_1, \hr_2)]_{\mathrm{BF}} =
\frac{i(-1)^{l+1}\,\sqrt{3}}{\sqrt{l(l+1)(2l+1)}}
P^1_l(\theta),
\end{equation}
where $P^1_l(\theta)$ is the associated Legendre polynomial,
$P^1_l(\theta)=-\sin\theta\, P'_l(\cos\theta)$.
As a result, the coefficient (\ref{eq:bh-h}) assumes the form
\begin{equation}
  \label{eq:h100}
  h^{(100)}_{2n} = 4\pi^3
\sum_{l=1}^n \frac{2l+1}{l(l+1)}
C^{(l, l)}_{n-l} (\alpha)\,C^{(l, l)}_{n-l} (\alpha')\,
P^1_l (\theta)\,P^1_l(\theta').
\end{equation}
Coefficients with odd index $n$ vanish, $h^{(100)}_{2n+1}=0$.


\section{Conclusion}
\label{sec:conclusion}

In the presented paper the perturbation theory was applied to the
calculation of the wave function of the three-body system in which the
potential $U$ of the inter-particle interaction is small comparing to the
kinetic energy.

The expression for the wave function in the zeroth- and first-order
approximation with respect to $U$ is given by eq~(\ref{eq:pt-psi-tot}).
The problem of the calculation of the matrix elements between the continuum
state wave function and the wave function having well-defined angular momentum
quantum numbers has been considered in Sec. \ref{sec:6d-hypersph}.
The calculation procedure leads to the appearance of twelve-dimensional
integrals (see eq.~(\ref{eq:me-2b})) which, in the most important particular
case of an isolated system, can be reduced to six-dimensional integrals, see
eq.~(\ref{eq:m1-1}).
The reduction was achieved by employing the technique of collective multipole
expansions in terms of Wigner $D$-functions (see eqs.~(\ref{eq:psi-i1}) and
(\ref{eq:i-1})) describing the rotation from the body-fixed frame (BF) to the
laboratory-fixed frame (LF).

The collective multipole coefficients (CMC) of the expansion of the Green
function of the system of three free particles were calculated in
Sec.~\ref{sec:mult-coeff-abf} and \ref{sec:hypersph-mult-coefs}.
The two different approaches were used for the calculation of CMC.
The first one (Sec.~\ref{sec:mult-coeff-abf}) is the straightforward
computation of the integral defining CMC using BF and LF corresponding to the
so-called angle-bisector gauge (see fig.~\ref{fig:abf}).
The expression for CMC in that approach is given by eq.~(\ref{eq:g-1}).

The second (hyperspherical) approach was based on the expansion of the
three-body free Green function over the basis set of six-dimensional 
hyperspherical harmonics.
Note that those were not conventional harmonics as defined in
\cite{Bateman-II,avery85jmp:_hyp_spher} but the set of harmonics labeled by
the individual angular momenta quantum numbers of particles
\cite{knirk74:_HSH_individual_ang_mom}.
The expression for CMC was then extracted from the series by using
some specific gauge convention (eqs.~(\ref{eq:g-hsh}), (\ref{eq:bh-h}) of
Sec.~\ref{sec:hypersph-mult-coefs}).

The advantage of the straightforward approach is that the resulting series
representation of CMC is simpler compared to that of the hyperspherical
approach.
Also, it remains valid in the whole shape space while the hyperspherical
series diverge at the configuration when $R=R'$.
Nevertheless, the hyperspherical representation can be more favorable if the
potential energy can be factorized in terms of hyperspherical coordinates.
In this case the many-dimensional integrals of the perturbation theory reduce
to the one-dimensional form.

Both representations of CMC are series of Hankel functions $H^{(1,2)}_n$
depending on hyperradial variables multiplied with weight functions depending
on the hyperangles.
These series converge rather well.
For example, at $Q=10$, $R=1$, $R=2$, $\alpha=\alpha'=\theta=\theta'=45^o$
the number of terms in series which are needed to achieve the accuracy of
$10^{-6}$ is about thirty.
The convergence rate decreases as $Q$ and/or $R$'s increase.
The same behavior is observed as the configuration triangles
(see fig.~\ref{fig:3bd-jacobi}) become narrower at the same $Q$ and $R$'s.

In general case of $j>0$, CMC are functions of \textit{six} variables,
i.e. two times the dimensionality of the shape space of three particles.
Surprisingly, from eq.~(\ref{eq:g-s}) it follows that CMC corresponding to
an $S$-state (with $j=0$) depends only on \textit{three} variables.
(Note that this property of $S$-state CMC is not seen from (\ref{eq:bh-s-1})
of the hyperspherical approach.)
Such property of $S$-state CMC was already noticed in
\cite{avm05:_rigid_d3h_jpb} where the collective multipole expansion of the
product of three-dimensional plane waves has been considered.
This fact still needs some physical explanation.

Currently, the work on the application of the developed theory to the problem
of the three-particle fragmentation process is in progress.

\section{Acknowledgments}
\label{sec:acknowledgments}

This work has been supported in part by the joint BRHE program of CRDF and
Russian Ministry of Education (grant No Y2-CP-10-02) and by the grant from the
``Dynasty foundation''.

\appendix

\section{The integration of $p^n$ over Euler angles}
\label{sec:integr-euler-angles}

In order to calculate the integral \eqref{eq:i-4} we employ the expression for
Wigner $D$-functions in terms of Euler angles,
$D^j_{\mu,\nu}(\Omega) = e^{-i(\alpha+\gamma)}\,d^j_{\mu,\nu}(\beta)$.
First, we calculate the integral over the angles $\alpha$ and $\gamma$,
\begin{equation}
  \label{eq:ai-1}
I_{n\mu\nu}= \int_0^{2\pi} \int_0^{2\pi} e^{i(\mu\alpha+\nu\gamma)}\, 
(a_1 \sin(\alpha+\gamma+\delta_1) - a_2 \sin(\alpha-\gamma+\delta_2))^n
\, d \alpha\, d \gamma,
\end{equation}
This integral can be calculated using the exponential representation of sine
functions.
It turns out that the integral is non-zero only for indices $n$, $\mu$, $\nu$
having the same parity.
Omitting details of routine transformations, we write
\begin{equation}
  \label{eq:ai-3}
 I_{n\mu\nu}= 4 \pi^2\, (-1)^{\frac{n+\nu}{2}}
e^{-i \bmu \delta_1} \,e^{-i \bnu \delta_2}\,
\frac{n!}{(2i)^n}
\sum_{k}
\frac{a_1^{n-\bnu-2k}\, a_2^{2k+\bnu}}{k!\, (k+\bnu)!\,
(\frac{n-\mu}{2}-k)!\, (\frac{n+\nu}{2} -k)!},
\end{equation}
where the sum runs over all non-negative values of $k$ at which factorials
remain finite.
Note that factorials and indices $\bmu,\bnu$ in eq.~(\ref{eq:ai-3}) are always
integer numbers (as was mentioned above),
\begin{equation}
  \label{eq:bmu-bnu}
  \bmu=\frac{\mu+\nu}{2}, \quad \bnu=\frac{\mu-\nu}{2}.
\end{equation}
Using the explicit form of the function $d^j_{\mu,\nu}(\beta)$ we can
integrate (\ref{eq:ai-3}) over $\beta$ term by term.
The partial integrals are
\begin{multline}
\label{eq:d-int}
\int_0^\pi a_1^{n-\bnu-2k}\, a_2^{2k+\bnu}\, d^j_{\mu,\nu}(\beta)\, \sin\beta
\, d \beta=
2 \sqrt{(j+\mu)!\,(j-\mu)!\, (j+\nu)!\,(j-\nu)!} \\
\times
\sum_{q} \rho_1^{n-\bnu-2k}\, \rho_2^{2k+\bnu}
\frac{(-1)^{j-\nu+q}\, (q+n+\nu-2k)!\, (j-q-\nu+2k)!}{q!\,(j-\mu-q)!\,
(j-\nu-q)!\, (\nu+\mu+q)!\, (j+n+1)!}.
\end{multline}
Substituting eq.~(\ref{eq:d-int}) into eq.~(\ref{eq:ai-3}) we can re-write the
summation over $k$ in (\ref{eq:ai-3}) as
\begin{multline}
  \label{eq:sum_k-1}
  \sum_{k} \left( \frac{\rho_2}{\rho_1}\right)^{2k}
\frac{(q+n+\nu-2k)!\, (j-q-\nu+2k)!}{k!\, (k+\bnu)!\,
(\frac{n-\mu}{2}-k)!\, (\frac{n+\nu}{2} -k)!}
=  2^{\frac{n-\nu}{2}}
 \sum_{k} \left( \frac{\rho_2}{\rho_1}\right)^{2k} \\
\times 
\frac{(2k+\mu-\nu-1)!!\, (n-\mu-2k-1)!!}{k!\,
  (\frac{n+\nu}{2}-k)!} 
\frac{(q+n+\nu-2k)!\, (j-q-\nu+2k)!}{(n-\mu-2k)!\, (\mu-\nu+2k)!}.
\end{multline}
The last multiplicand on rhs on this equation can be presented in a compact
differential form
\begin{equation}
  \label{eq:ratio-diff}
  \frac{(q+n+\nu-2k)!\, (j-q-\nu+2k)!}{(n-\mu-2k)!\, (\mu-\nu+2k)!}
= (-1)^{q+\mu+\nu} \,\deriv^{j-\mu-q} t^{n+j+1} \deriv^{\mu+\nu+q}
t^{-n+\mu+2k-1} \bigr|_{t=1}.
\end{equation}
Substituting this identity into (\ref{eq:sum_k-1}) we put derivatives outside
the summation.
Thereby, it evaluates in closed form
\begin{multline}
  \label{eq:sum_k-2}
  \sum_{k} (\ldots) =  (-1)^{\mu+\nu+q}\, 2^{\frac{n-\nu}{2}}
\frac{(\mu-\nu-1)!!\, (n-\mu-1)!!}{(\frac{n+\nu}{2})!}\\
\times
\deriv^{j-\mu-q} t^{n+j+1} \deriv^{\mu+\nu+q}
t^{\mu-n-1}\,
{}_2F_1 \left(
\frac{1+\mu-\nu}{2}, \, - \frac{n+\nu}{2};\, \frac{1+\mu-n}{2};\,
\frac{(t \rho_2)^2}{\rho_1^2}
\right) \biggr|_{t=1},
\end{multline}
where ${}_2F_1$ denotes Gauss hypergeometric function \cite{Bateman-I}.

The final step is to rearrange the exterior summation over $q$,
\begin{multline}
  \label{eq:sum_q-1}
  \sum_{q} \frac{1}{q!\,(j-\mu-q)!\,(j-\nu-q)!\,(\nu+\mu+q)!}
\deriv^{j-\mu-q} t^{n+j+1} \deriv^{q+\mu+\nu} 
=  t^{n+j+1-m} \\
\times \sum_{m} \frac{(n+j+1)!}{m!\, (n+j+1-m)!}   \deriv^{j+\nu-m}
\sum_{q} \frac{1}{q!\,(j-\nu-q)!\,(\nu+\mu+q)!\,(j-\mu-q-m)!}.
\end{multline}
Here, the decouple sum over $q$ evaluates to a closed form,
\begin{equation}
  \label{eq:sum_q-2}
\sum_{q} (\ldots) = \frac{(2j-m)!}{(j+\mu)!\,(j-\nu)!\,(j-\mu-m)!\,
(j+\nu-m)!}.
\end{equation}

Collecting above eqs.~(\ref{eq:ai-3}), (\ref{eq:d-int}), (\ref{eq:sum_k-2}) --
(\ref{eq:sum_q-2}) we finally write the expression for the integral
\eqref{eq:i-4},
\begin{multline}
  \label{eq:int-pn}
P_n^{(j\mu\nu)} = 8 \pi^2\,
i^n (-1)^{j+\frac{n+\nu}{2}} \rho_1^n\,
e^{-i \delta_1(\mu+\nu)/2} \,
\left( \frac{\rho_2}{\rho_1} e^{-i \delta_2} \right)^{(\mu-\nu)/2}
\frac{(\mu-\nu-1)!!\, (n-\mu-1)!!}{(n+\nu)!!} \\
\times
\sqrt\frac{(j-\mu)!\,(j+\nu)!}{(j+\mu)!\,(j-\nu)!}
\sum_{m} \frac{(2j-m)!}{m!\, (j-\mu-m)!\, (j+\nu-m)!\, (n+j+1-m)!} \\
\times
 \deriv^{j+\nu-m} t^{\mu-n-1}\,
{}_2F_1 \left(
\frac{1+\mu-\nu}{2}, \, - \frac{n+\nu}{2};\, \frac{1+\mu-n}{2};\,
\frac{(t \rho_2)^2}{\rho_1^2}
\right) \biggr|_{t=1}.
\end{multline}
Note that the derivative can be written as a combination of Gauss
hypergeometric functions,
\begin{multline}
  \label{eq:f21-d-1}
\deriv^{j+\nu-m} t^{\mu-n-1} \, {}_2F_1 \left(
\frac{1+\mu-\nu}{2}, \, - \frac{n+\nu}{2};\, \frac{1+\mu-n}{2};\,
\frac{(t \rho_2)^2}{\rho_1^2}
\right) \biggr|_{t=1} \\
=\sum_{\substack{\alpha = (j+\nu-m),\\ (j+\nu-m)-1, \ldots}}
\frac{(-1)^\alpha\, 2^{2\alpha-j-\nu+m}\,(j+\nu-m)!}{(2\alpha-j-\nu+m)!\,
(j+\nu-m-\alpha)!}  \\
\times
\left( \frac{1-\mu+n}{2} \right)_\alpha\,
{}_2F_1\left(
\frac{1+\mu-\nu}{2}, \, - \frac{n+\nu}{2};\, \frac{1+\mu-n}{2}-\alpha;\,
\frac{\rho_2^2}{\rho_1^2}
\right),
\end{multline}
where $(\frac{1-\mu+n}{2})_\alpha$ denotes Pochhammer symbol
\cite{Bateman-I}.

Using eq.~\eqref{eq:int-pn}, one can easily calculate the coefficient
$P_{2n}^{(000)}$ corresponding to an $S$-state,
\begin{equation}
  \label{eq:g000}
      P^{(000)}_{2n} = 8\pi^2\, \rho_1^{2n}\, 
\frac{(2n-1)!!}{(2n)!!\,(2n+1)!}\,
{}_2F_1 \left( 
\frac{1}{2},\, -n;\, \frac{1}{2}-n;\, \frac{\rho_2^2}{\rho_1^2}
\right).
\end{equation}
Here, the hypergeometric function reduces to the Legendre polynomial which
leads to eq.~(\ref{eq:g-s}).
For $P^e$-state with pseudotensor parity the non-zero parameter is
\begin{equation}
  \label{eq:g100}
P_{2n}^{(100)} =  -16 \pi^2\,
\frac{(\rho_1 \rho_2)^n}{(2n+2)!}\,
\left(
\deriv  + (n+1)
\right)\, t^{-n-1}\,
P_n \left( \frac{\rho_1^2 + t^2 \rho_2^2}{2 t \rho_1 \rho_2} \right)
 \Biggr|_{t=1}.
\end{equation}
Calculating the derivative in this identity we arrive at eq.~(\ref{eq:g-p}) of
the main text.
Note that both above parameters with odd index vanish, i.e.
$P^{(000)}_{2n+1}=P^{(100)}_{2n+1}=0$.


\section{The six-dimensional hyperspherical harmonics and 
their properties} 
\label{sec:hypersph-decomp-c2}

For the sake of completeness below we summarize the most important properties
of the set of six-dimensional hyperspherical harmonics which are
eigenfunctions of the particle's angular momentum operators.
The derivation of the results can be found, e.g. in
\cite{knirk74:_HSH_individual_ang_mom}.

The scalar product of two hyperspherical harmonics is defined by
\begin{multline}
  \label{eq:def-y6d-3}
  (Y_n (\hcr) \cdot Y_n (\hcr') ) 
= \sum_{l=0}^n \sum_{q=0}^{q_{max}} 
\sum_{mm'} Y^*_{nqlmm'} (\hcr)\, Y_{nqlmm'} (\hcr') 
= \sum_{ql} \frac{(2l+1)}{4\pi}   \\
\times (2[n-2q-l]+1) \,
 C^{(n-2q-l,\,l)}_q (\alpha)\,
C^{(n-2q-l,\,l)}_q (\alpha')\, P_l (\hr_1 \cdot \hr'_1)\, 
P_{N-2q-l} (\hr_2\cdot\hr'_2),
\end{multline}
where $q_{max}$ is the integer part of the ratio $(n-l)/2$ and the hyperangle
$\alpha$ is defined in (\ref{eq:def-hsc}).
The functions $C^{(n-2q-l,\, l)}_q$ are defined as
\begin{multline}
  \label{eq:def-cpol}
      C^{(l', l)}_{q} (\alpha) =
\sqrt\frac{2^{l+l'+3}\, (2q +l+l'+2)\,q!\, (q+l+l'+1)!}{ \pi\,
(2q+2l+1)!!\, (2q+2l'+1)!!}\\
\times
2^q \, (\sin\alpha)^{l'}\, (\cos\alpha)^l \,
P^{(l'+1/2,\, l+1/2)}_q (\cos 2\alpha),
\end{multline}
where $P^{(a,b)}_q$ is Jacobi polynomial \cite{Bateman-II}.
Above functions are orthonormal,
\begin{equation}
  \label{eq:orth-cpol}
  \int_0^{\pi/2} C^{(l', l)}_q(\alpha) \, 
C^{(l', l)}_{q'} (\alpha)\, 
\frac{\sin^2 2\alpha}{4}\, d\alpha = \delta_{q,q'}.
\end{equation}
The hyperspherical harmonics defined by eq.~(\ref{eq:def-y6d}) are
orthonormal,
\begin{equation}
  \label{eq:def-y6d-2}
  \int Y^*_{nqlmm'} (\hcr) \, Y^*_{nqlmm'} (\hcr) \,
d^5 \Omega = d_{n,n_1} d_{q,q_1} d_{l,l_1} d_{m,m_1} d_{m',m'_1},
\end{equation}
where $d^5 \Omega$ is the surface elements of the six-dimensional hypersphere, 
which can be written as
\begin{equation}
  \label{eq:ve-2}
d^5 \Omega = \frac{\sin^2 2\alpha}{4} \,
\sin\theta_1\, \sin\theta_2\, d\alpha\, d\theta_1\, d\theta_2\,
d\phi_1 \, d \phi_2,
\end{equation}
where $\theta_{1,2},\phi_{1,2}$ are three-dimensional spherical angles of the
unit vectors $\hr_{1,2}$.
Note that the total solid angle in the six-dimensional space is
\begin{equation}
  \label{eq:d6-om}
  \int_{S_6} d^5 \Omega = \pi^3,
\end{equation}
where $S_6$ is the six-dimensional hypersphere of the unit radius.


\begin{thebibliography}{16}
\expandafter\ifx\csname natexlab\endcsname\relax\def\natexlab#1{#1}\fi
\expandafter\ifx\csname bibnamefont\endcsname\relax
  \def\bibnamefont#1{#1}\fi
\expandafter\ifx\csname bibfnamefont\endcsname\relax
  \def\bibfnamefont#1{#1}\fi
\expandafter\ifx\csname citenamefont\endcsname\relax
  \def\citenamefont#1{#1}\fi
\expandafter\ifx\csname url\endcsname\relax
  \def\url#1{\texttt{#1}}\fi
\expandafter\ifx\csname urlprefix\endcsname\relax\def\urlprefix{URL }\fi
\providecommand{\bibinfo}[2]{#2}
\providecommand{\eprint}[2][]{\url{#2}}

\bibitem[{\citenamefont{Briggs and Schmidt}(2000)}]{briggs00}
\bibinfo{author}{\bibfnamefont{J.~S.} \bibnamefont{Briggs}} \bibnamefont{and}
  \bibinfo{author}{\bibfnamefont{V.}~\bibnamefont{Schmidt}},
  \bibinfo{journal}{J. Phys. B: At. Mol. Opt. Phys.}
  \textbf{\bibinfo{volume}{33}}, \bibinfo{pages}{R1} (\bibinfo{year}{2000}).

\bibitem[{\citenamefont{Datz et~al.}(2000)\citenamefont{Datz, Thomas, Rosen,
  Larsson, Derkatch, Hellberg, and van~der
  Zande}}]{datz00:_diss_recomb_water_angl_distr}
\bibinfo{author}{\bibfnamefont{S.}~\bibnamefont{Datz}},
  \bibinfo{author}{\bibfnamefont{R.}~\bibnamefont{Thomas}},
  \bibinfo{author}{\bibfnamefont{S.}~\bibnamefont{Rosen}},
  \bibinfo{author}{\bibfnamefont{M.}~\bibnamefont{Larsson}},
  \bibinfo{author}{\bibfnamefont{A.~M.} \bibnamefont{Derkatch}},
  \bibinfo{author}{\bibfnamefont{F.}~\bibnamefont{Hellberg}}, \bibnamefont{and}
  \bibinfo{author}{\bibfnamefont{W.}~\bibnamefont{van~der Zande}},
  \bibinfo{journal}{Phys. Rev. Lett.} \textbf{\bibinfo{volume}{85}},
  \bibinfo{pages}{5555} (\bibinfo{year}{2000}).

\bibitem[{\citenamefont{M\"uller et~al.}(1999)\citenamefont{M\"uller, Eckert,
  Braun, and Helm}}]{helm99:_h3_first}
\bibinfo{author}{\bibfnamefont{U.}~\bibnamefont{M\"uller}},
  \bibinfo{author}{\bibfnamefont{T.}~\bibnamefont{Eckert}},
  \bibinfo{author}{\bibfnamefont{M.}~\bibnamefont{Braun}}, \bibnamefont{and}
  \bibinfo{author}{\bibfnamefont{H.}~\bibnamefont{Helm}},
  \bibinfo{journal}{Phys. Rev. Lett.} \textbf{\bibinfo{volume}{83}},
  \bibinfo{pages}{2718} (\bibinfo{year}{1999}).

\bibitem[{\citenamefont{Galster et~al.}(2004)\citenamefont{Galster, M\"uller,
  and Helm}}]{galster04:_h3_prl}
\bibinfo{author}{\bibfnamefont{U.}~\bibnamefont{Galster}},
  \bibinfo{author}{\bibfnamefont{U.}~\bibnamefont{M\"uller}}, \bibnamefont{and}
  \bibinfo{author}{\bibfnamefont{H.}~\bibnamefont{Helm}},
  \bibinfo{journal}{Phys. Rev. Lett.} \textbf{\bibinfo{volume}{92}},
  \bibinfo{pages}{073002} (\bibinfo{year}{2004}).

\bibitem[{\citenamefont{Galster et~al.}(2005)\citenamefont{Galster,
  Baumgartner, M\"uller, Helm, and Jungen}}]{galster_helm05:_h3_results}
\bibinfo{author}{\bibfnamefont{U.}~\bibnamefont{Galster}},
  \bibinfo{author}{\bibfnamefont{F.}~\bibnamefont{Baumgartner}},
  \bibinfo{author}{\bibfnamefont{U.}~\bibnamefont{M\"uller}},
  \bibinfo{author}{\bibfnamefont{H.}~\bibnamefont{Helm}}, \bibnamefont{and}
  \bibinfo{author}{\bibfnamefont{M.}~\bibnamefont{Jungen}},
  \bibinfo{journal}{Phys. Rev. A} \textbf{\bibinfo{volume}{72}},
  \bibinfo{pages}{062506} (\bibinfo{year}{2005}).

\bibitem[{\citenamefont{Faddeev}(1960)}]{Faddeev_eqs60}
\bibinfo{author}{\bibfnamefont{L.}~\bibnamefont{Faddeev}},
  \bibinfo{journal}{Zh. Eksperim. i Teor. Fiz.} \textbf{\bibinfo{volume}{39}},
  \bibinfo{pages}{1459} (\bibinfo{year}{1960}), \bibinfo{note}{[{Soviet Phys.}
  -- JETP, \textbf{12}, p.1014, 1961]}.

\bibitem[{\citenamefont{Meremianin}(2005)}]{avm05:_rigid_d3h_jpb}
\bibinfo{author}{\bibfnamefont{A.~V.} \bibnamefont{Meremianin}},
  \bibinfo{journal}{J. Phys. B: At. Mol. Opt. Phys.}
  \textbf{\bibinfo{volume}{38}}, \bibinfo{pages}{757} (\bibinfo{year}{2005}).

\bibitem[{\citenamefont{Erdelyi
  et~al.}(1953{\natexlab{a}})\citenamefont{Erdelyi, Magnus, Oberhettinger, and
  Tricomi}}]{Bateman-II}
\bibinfo{author}{\bibfnamefont{A.}~\bibnamefont{Erdelyi}},
  \bibinfo{author}{\bibfnamefont{W.}~\bibnamefont{Magnus}},
  \bibinfo{author}{\bibfnamefont{F.}~\bibnamefont{Oberhettinger}},
  \bibnamefont{and} \bibinfo{author}{\bibfnamefont{F.~G.}
  \bibnamefont{Tricomi}}, \emph{\bibinfo{title}{Higher trancendental functions.
  Bateman manuscript project}}, vol.~\bibinfo{volume}{II}
  (\bibinfo{publisher}{McGraw-hill book company, Inc},
  \bibinfo{year}{1953}{\natexlab{a}}).

\bibitem[{\citenamefont{Varshalovich et~al.}(1988)\citenamefont{Varshalovich,
  Moskalev, and Khersonskii}}]{Varsh}
\bibinfo{author}{\bibfnamefont{D.~A.} \bibnamefont{Varshalovich}},
  \bibinfo{author}{\bibfnamefont{A.~N.} \bibnamefont{Moskalev}},
  \bibnamefont{and} \bibinfo{author}{\bibfnamefont{V.~K.}
  \bibnamefont{Khersonskii}}, \emph{\bibinfo{title}{Quantum theory of angular
  momentum}} (\bibinfo{publisher}{World Scientific},
  \bibinfo{address}{Singapore}, \bibinfo{year}{1988}).

\bibitem[{\citenamefont{Littlejohn and Reinsch}(1997)}]{littlejohn97:_gauge}
\bibinfo{author}{\bibfnamefont{R.~G.} \bibnamefont{Littlejohn}}
  \bibnamefont{and} \bibinfo{author}{\bibfnamefont{M.}~\bibnamefont{Reinsch}},
  \bibinfo{journal}{Rev. Mod. Phys.} \textbf{\bibinfo{volume}{69}},
  \bibinfo{pages}{213} (\bibinfo{year}{1997}).

\bibitem[{\citenamefont{Meremianin and Briggs}(2003)}]{meremianin03:_phys_rep}
\bibinfo{author}{\bibfnamefont{A.~V.} \bibnamefont{Meremianin}}
  \bibnamefont{and} \bibinfo{author}{\bibfnamefont{J.~S.}
  \bibnamefont{Briggs}}, \bibinfo{journal}{Phys. Rep.}
  \textbf{\bibinfo{volume}{384}}, \bibinfo{pages}{121} (\bibinfo{year}{2003}).

\bibitem[{\citenamefont{Wen and Avery}(1985)}]{avery85jmp:_hyp_spher}
\bibinfo{author}{\bibfnamefont{Z.-Y.} \bibnamefont{Wen}} \bibnamefont{and}
  \bibinfo{author}{\bibfnamefont{J.}~\bibnamefont{Avery}}, \bibinfo{journal}{J.
  Math. Phys.} \textbf{\bibinfo{volume}{26}}, \bibinfo{pages}{396}
  (\bibinfo{year}{1985}).

\bibitem[{\citenamefont{Manakov et~al.}(1996)\citenamefont{Manakov, Marmo, and
  Meremianin}}]{Man-96}
\bibinfo{author}{\bibfnamefont{N.~L.} \bibnamefont{Manakov}},
  \bibinfo{author}{\bibfnamefont{S.~I.} \bibnamefont{Marmo}}, \bibnamefont{and}
  \bibinfo{author}{\bibfnamefont{A.~V.} \bibnamefont{Meremianin}},
  \bibinfo{journal}{J. Phys. B: At. Mol. Opt. Phys.}
  \textbf{\bibinfo{volume}{29}}, \bibinfo{pages}{2711} (\bibinfo{year}{1996}).

\bibitem[{\citenamefont{Knirk}(1974)}]{knirk74:_HSH_individual_ang_mom}
\bibinfo{author}{\bibfnamefont{D.~L.} \bibnamefont{Knirk}},
  \bibinfo{journal}{J. Chem. Phys.} \textbf{\bibinfo{volume}{60}},
  \bibinfo{pages}{66} (\bibinfo{year}{1974}).

\bibitem[{\citenamefont{Erdelyi
  et~al.}(1953{\natexlab{b}})\citenamefont{Erdelyi, Magnus, Oberhettinger, and
  Tricomi}}]{Bateman-I}
\bibinfo{author}{\bibfnamefont{A.}~\bibnamefont{Erdelyi}},
  \bibinfo{author}{\bibfnamefont{W.}~\bibnamefont{Magnus}},
  \bibinfo{author}{\bibfnamefont{F.}~\bibnamefont{Oberhettinger}},
  \bibnamefont{and} \bibinfo{author}{\bibfnamefont{F.~G.}
  \bibnamefont{Tricomi}}, \emph{\bibinfo{title}{Higher transcendental
  functions}}, vol.~\bibinfo{volume}{I} (\bibinfo{publisher}{McGraw-hill book
  company, Inc}, \bibinfo{year}{1953}{\natexlab{b}}).

\bibitem[{\citenamefont{Landau and Lifshitz}(1977)}]{landau3tom_eng}
\bibinfo{author}{\bibfnamefont{L.~D.} \bibnamefont{Landau}} \bibnamefont{and}
  \bibinfo{author}{\bibfnamefont{E.~M.} \bibnamefont{Lifshitz}},
  \emph{\bibinfo{title}{Quantum mechanics}} (\bibinfo{publisher}{Pergamon},
  \bibinfo{address}{New-York}, \bibinfo{year}{1977}).

\end{thebibliography}

\end{document}